\documentclass[twocolumn,showpacs,preprintnumbers,amsmath,amssymb]{revtex4}
\usepackage{epsfig}
\usepackage{dcolumn}% Align table columns on decimal point
\usepackage{bm}% bold math
\newcommand{\bn}{\bm{n}}

\newcommand{\be}{\begin{equation}}
\newcommand{\ee}{\end{equation}}
\newcommand{\bea}{\begin{eqnarray}}
\newcommand{\eea}{\end{eqnarray}}

\newcommand{\btab}{\begin{tabbing}}
\newcommand{\etab}{\end{tabbing}}

\newcommand{\bit}{\begin{itemize}}
\newcommand{\eit}{\end{itemize}}

\newcommand{\ben}{\begin{enumerate}}
\newcommand{\een}{\end{enumerate}}

\newcommand{\bfig}{\begin{figure}[h]}
\newcommand{\efig}{\end{figure}}

\newcommand{\hs}[1]{\hspace*{#1}}

\newcommand{\LB}[1]{\label{#1}}
\begin{document}

\title{CGL Defect Chaos and Bursts in Hexagonal Rotating
non-Boussinesq Convection}
\author{Santiago Madruga$^+$, Hermann Riecke$^+$, and Werner Pesch$^*$}
\affiliation{$^+$Engineering Science and Applied Mathematics, 
Northwestern University, Evanston, IL 60208, USA, $^*$Physikalisches Institut,
Universit\"at Bayreuth, D-95440 Bayreuth, Germany}
\date{\today}

\begin{abstract}
 
We employ numerical computations of the full Navier-Stokes equations
to investigate non-Boussinesq convection in a rotating system using
water as a working fluid. We identify two regimes. For weak
non-Boussinesq effects the Hopf bifurcation from steady to oscillating
(whirling) hexagons is supercritical and typical states exhibit defect
chaos that is systematically described by the cubic complex
Ginzburg-Landau equation. For stronger non-Boussinesq effects the Hopf
bifurcation becomes subcritical and the oscillations exhibit localized
chaotic bursting, which can be modeled by a quintic complex
Ginzburg-Landau equation.

\end{abstract}  
\maketitle

The complex Ginzburg-Landau equation (CGL) as the universal
description of weakly nonlinear oscillations has been studied
theoretically in great detail.  The classical, supercritical
case, which involves only two independent parameters, looks
deceptively simple. It exhibits, however, a vast variety of
qualitatively different phenomena including different types of
spatio-temporally chaos (e.g. \cite{ChMa96,ArKr02}). If
the oscillations arise in a subcritical Hopf bifurcation the
quintic CGL comes into play, which introduces further
interesting states including, for instance, intermittent bursts
in the oscillation amplitude.

In contrast to the extensive theoretical work on the CGL, direct
experimental validation of its various regimes of complex behavior are
rather scarce, in particular for the two-dimensional case
\cite{OuFl96}. In this paper we present detailed numerical
computations for an experimentally realizable thermal convection
experiment that exhibit defect chaos and bursts. We show that for the
defect chaos the cubic CGL should provide a systematic description,
while the bursts can be modeled with a quintic CGL.

Rayleigh-B\'enard convection of a fluid layer heated from below
in systems with large aspect ratio, in which the lateral
dimension $L$ of the layer is much larger than its thickness
$h$, has proved to be a paradigmatic experimental system
\cite{BoPe00} for studies of complex patterns. Above a critical
temperature difference $\Delta T_c$ across the layer, which
corresponds to the critical value $R_c$ of the dimensionless
Rayleigh number $R$, one observes in the simplest case the
familiar striped (roll) patterns with wavenumbers $q$ close to
the critical wavenumber $q_c$. However, in systems in which 
$\Delta T_c$ is large fluid properties like the thermal
expansion coefficient or the viscosity vary significantly across
the layer. Under these non-Boussinesq (NB) conditions the
instability of the homogeneous state leads to hexagonal
convection patterns \cite{Bu67}. 

\begin{figure}
\epsfxsize=4.6cm {\epsfbox{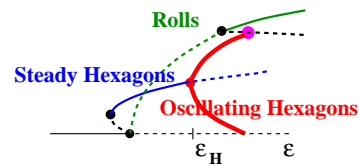}}
\caption{Sketch of bifurcation diagram for rotating NB-convection.
Solid (dashed) lines denote stable (unstable) branches. For the
oscillating (`whirling') hexagons maxima and minima of the
oscillations are indicated (thick solid lines).
}
\LB{f:bifdia}
\end{figure}

If the chiral symmetry of the system is broken by rotating the
layer about a vertical axis with frequency $\Omega$, new
interesting dynamics arise. For $\Omega$ above the
K\"uppers-Lortz frequency $\Omega_{KL}$ one finds in the
Boussinesq case immediately at onset $R_c$ domain chaos in which
rolls persistently switch orientation (cf. \cite{BoPe00}). In
contrast, in the NB case the hexagons are steady in this regime.
However, weakly nonlinear theory predicts a secondary
oscillatory instability at $\epsilon\equiv
(R-R_c)/R_c=\epsilon_{H}$ that leads to `whirling hexagons' in
which the three hexagon amplitudes oscillate about their mean,
with their phases shifted by $2\pi/3$ relative to each other
\cite{Sw84,So85,EcRi00a,EcRi00b} (cf. Fig.\ref{f:bifdia}).

\begin{figure} 
\epsfxsize=4.2cm {\epsfbox{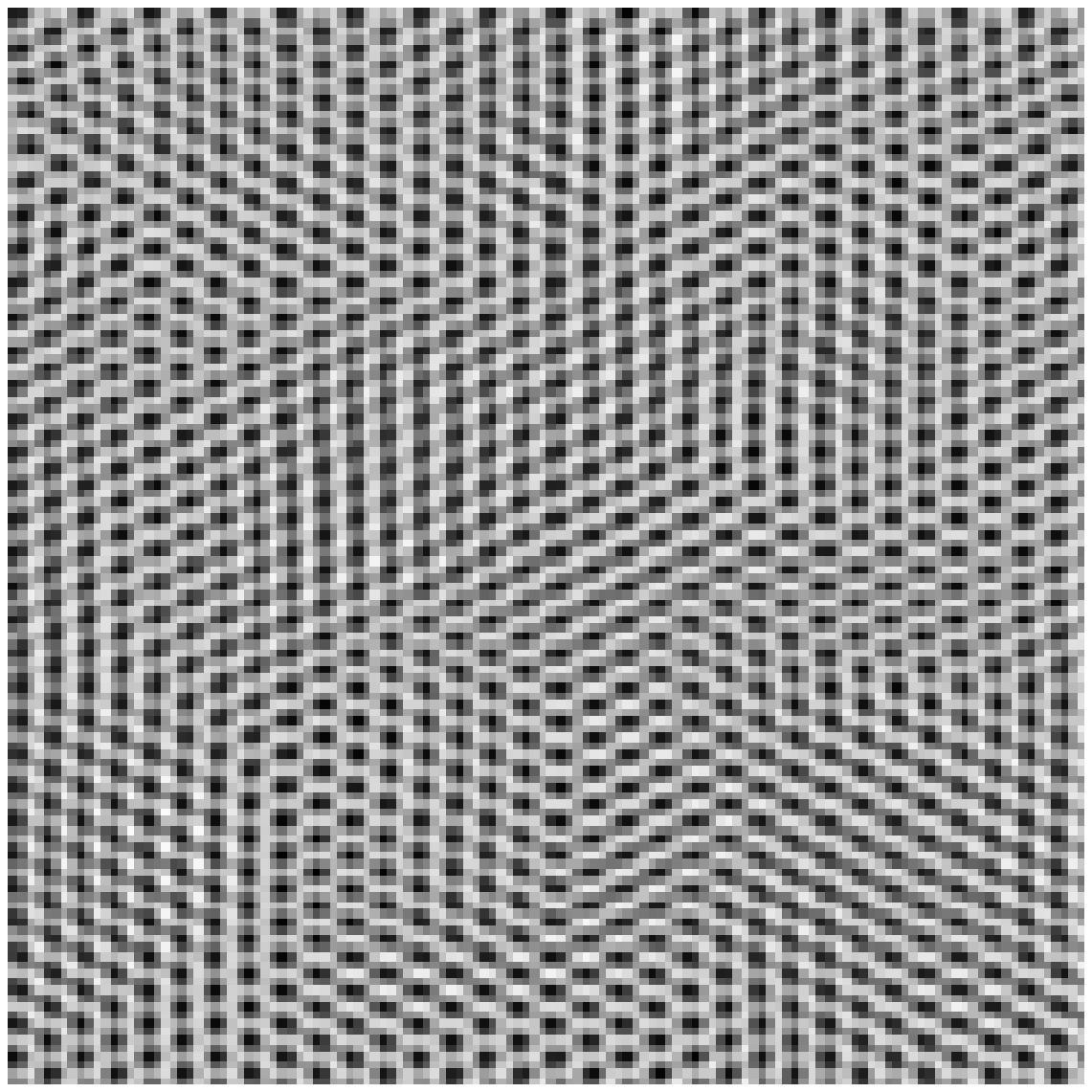}} 
\epsfxsize=4.2cm {\epsfbox{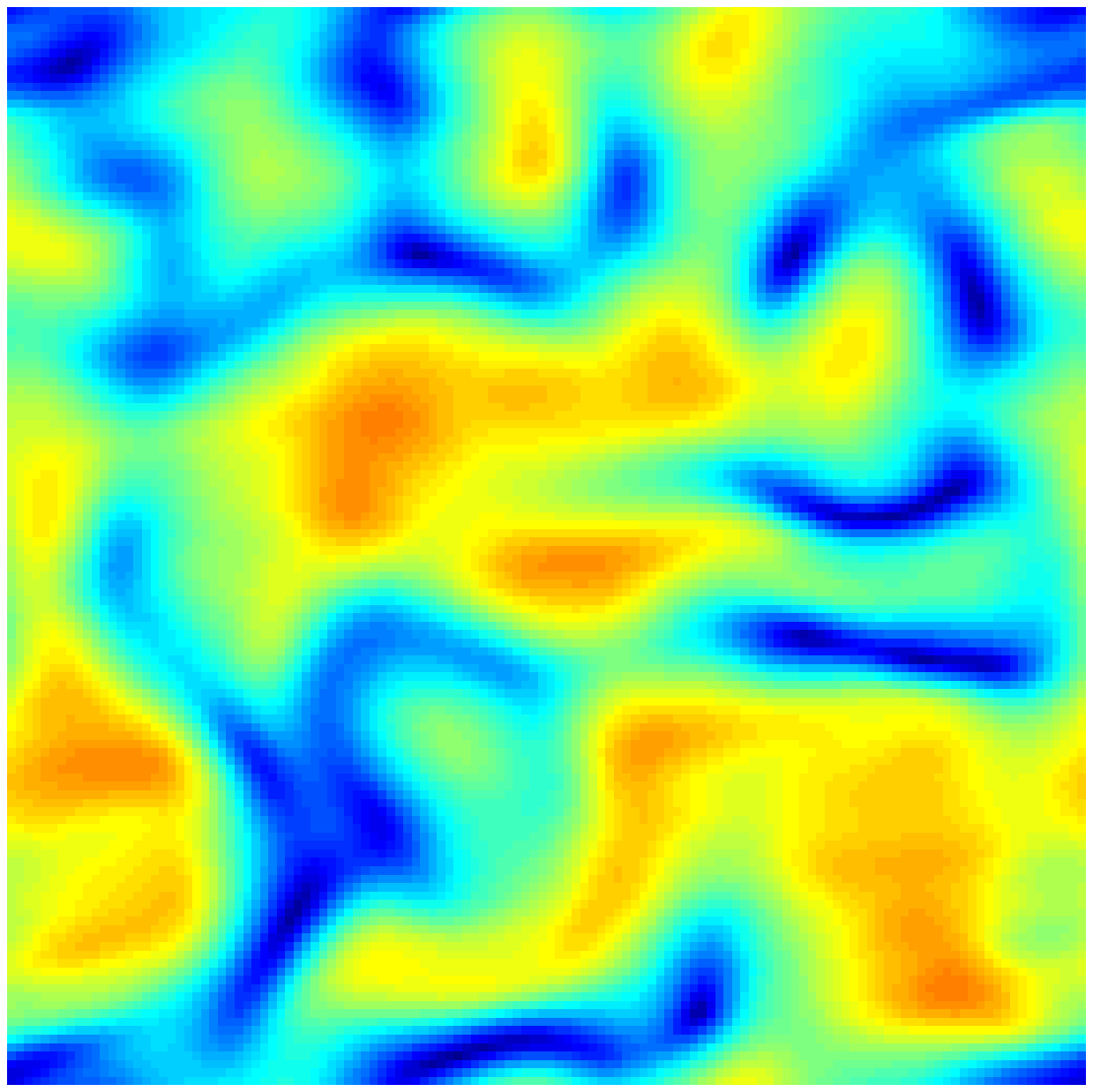}} 
\caption{Disordered state of whirling hexagons in case $A$ 
for $\epsilon=0.2$ ($R=7396.6$), $L=32\cdot 2\pi/q_c=38.3$.
a) Snapshot of $\theta({\bf r})$. 
b) Magnitude $|{\mathcal H}|$ of oscillation amplitude of snapshot shown in a). 
Red (blue) indicates large (small) values of $|{\mathcal H}|$.}
\LB{f:dis-super} \end{figure} 

Here we perform computer experiments on rotating non-Boussinesq
convection with water as working fluid. The experimental
realization of these simulations should pose no serious
difficulty in presently available experimental setups. Focusing
on the whirling hexagons, we obtain different spatio-temporally
disordered states, an example of which is shown in
Fig.\ref{f:dis-super}a, and use the complex oscillation
amplitude ${\mathcal H}$ (see (\ref{e:demod}) below) to
interpret them in terms of suitable complex Ginzburg-Landau
equations. The magnitude $|{\mathcal H}|$ is shown in
Fig.\ref{f:dis-super}b.

Specifically, we consider two situations corresponding to weak
and intermediate NB effects, respectively, and solve the full
Navier-Stokes (NS) equations with leading-order NB effects
included, as discussed previously \cite{YoRi03b,MaRi05}. We
obtain spatially periodic stationary hexagon patterns and test
their stability using a standard Galerkin method. The temporal
evolution of the states is then simulated with a pseudo-spectral
code with periodic lateral boundary conditions
\cite{YoRi03b,MaRi05}. We consider two set-ups differing in the
mean temperature $T_0$ of the fluid layer. In \underline{case $A$} we choose
$T_0=14^\circ {\rm C}$ resulting in $\Delta T_c=6.4^\circ {\rm
C}$  and  a value of  $Q=-2.2$ for Busse's NB parameter
\cite{Bu67}; \underline{case $B$}  with $T_0=12^\circ {\rm C}$  and $\Delta
T_c=8.3^\circ {\rm C}$ is more non-Boussinesq and yields
$Q=-3.62$ \footnote{The NB coefficients (cf.
\cite{YoRi03b,MaRi05}) at threshold are given by
$\gamma_0^{(c)}=0.0009$, $\gamma_1^{(c)}=0.2937$,
$\gamma_2^{(c)}=-0.1681$, $\gamma_3^{(c)}=0.0215$,
$\gamma_4^{(c)}=-0.0022$  for case $A$ and
$\gamma_0^{(c)}=0.0010$, $\gamma_1^{(c)}=0.4885$,
$\gamma_2^{(c)}=-0.2281$, $\gamma_3^{(c)}=0.0287$,
$\gamma_4^{(c)}=-0.0032$ for case $B$.}. In both cases
$h=0.492 \,{\rm cm}$ and  $\Omega=65 \, \nu_0/h^2$ with $\nu_0$ being 
the viscosity of the fluid in the
mid-plane.   

In \underline{case $A$} we obtain above the Hopf bifurcation at $\epsilon_H$
states in which almost all convection cells are oscillating about a
mean, but for general initial conditions the oscillations of different
cells are out of phase with respect to each other. This is shown in
Fig.\ref{f:dis-super}a where the solution is visualized by the
deviation $\theta({\bf r},t)$ of the temperature from the conduction
profile in the mid-plane of the fluid layer. Here ${\bf r}$
denotes the horizontal coordinates. To extract the oscillations
explicitly we make use of the fact that the underlying hexagon pattern
itself is well ordered. We first demodulate each snapshot in space by
writing  
\bea 
{ \theta}({\bf r},t)= \sum_{j=1}^3  A_j({\bf r},t)\, 
e^{i q_c \hat{\bn}_j\cdot {\bf r}}+c.c.+h.o.t.  
\eea  
Here the
wavevectors $q_c\hat{\bn}_j$ represent the three dominant  wavevectors
of the hexagons and $h.o.t$. denotes their harmonics.  For strictly
periodic stationary hexagon patterns the amplitudes $A_j$, $j =1..3$,
do not depend on ${\bf r}$ or $t$ and have the same modulus $A_{hex}$.
They can be chosen real and positive in our case (water). More
generally, the amplitudes $A_j({\bf r},t)$  depend slowly on space
representing the contributions from the side-bands of the basic
wavevectors $q_c\hat{\bn}_j$.

Slightly above the Hopf bifurcation the amplitudes $A_j$ vary in
time. For patterns that are close to periodic in space they can
be expressed as
\bea
A_j=\left(A_{hex}  +\mu^{1/2}\left[e^{j\,2\pi i/3} {\mathcal
H}\,e^{i\omega t}+c.c.\right]+{\cal O}(\mu)\right)\times
\nonumber \\
\exp\left(i\Delta q \,\hat{\bn}_j \cdot {\bf r}+
i\mu^{1/2}\phi_j({\bf r},t)\right), \hspace*{.2cm}
\LB{e:demod}
\eea
where $\mu = (R -R_{H})/R_{H}$ and $\omega$ is the Hopf frequency.
Here ${\mathcal H}$ denotes the complex oscillation amplitude, which
we extract by standard demodulation of a series of snapshots like
Fig.\ref{f:dis-super}a in space and time. The space
dependence of the phases $\phi_j$ captures slight deformations of the
underlying hexagon lattice. 

A representative  snapshot of the magnitude of the oscillation
amplitude is shown in Fig.\ref{f:dis-super}b. A more detailed analysis
shows that the domains with low oscillation magnitude are due to
spiral defects in the complex oscillation amplitude, i.e. ${\mathcal
H}=0$ at isolated points. During the evolution of the system these
defects are persistently created and annihilated in pairs. In
appearance, this state is thus quite similar to the defect chaos of
the cubic CGL (see below).

\begin{figure}
\epsfxsize=4.cm {\epsfbox{bifdia.Hopfback.eps}}
\epsfxsize=4.cm {\epsfbox{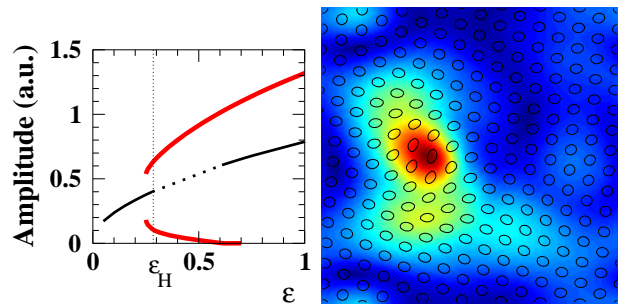}}
\caption{a) Bifurcation diagram for the subcritical case $B$
(cf. Fig.\ref{f:bifdia} with the rolls omitted).
b) Bursting state obtained with NS-simulation in case $B$
for $\epsilon=0.5$ ($R=6121.9$) and $L=16\cdot 2\pi/q_c=19.1$. 
$|{\mathcal H}|$ color-coded as in Fig.\ref{f:dis-super}b, 
contour lines indicate hexagon pattern.
}
\LB{f:snap-burst-NS}
\end{figure}

When the NB effects are stronger (\underline{case $B$}) the Hopf bifurcation is
shifted to larger values of $\epsilon$ and, unexpectedly, becomes
subcritical. The numerically determined bifurcation diagram showing
the jump in the oscillation amplitude, the small hysteresis, and the
restabilization (reentrance \cite{MaRi05}) of the steady hexagons is
presented in Fig.\ref{f:snap-burst-NS}a. As a consequence, the
dynamical states obtained in this regime are very different, as shown
in Fig.\ref{f:snap-burst-NS}b. The contour lines indicate the
underlying hexagon pattern and, as in Fig.\ref{f:dis-super}b, the
colors represent the oscillation magnitude. Strikingly, the
oscillations are now localized into relatively small domains. Only
within these domains the convection cells are elongated to footballs
and are whirling. In most parts of the system the hexagons are
relatively steady. The burst-like temporal evolution of the hexagon
amplitudes is evident from Fig.\ref{f:burst-t}. It shows the
normalized oscillation intensity in the bursts, which we define as
$I(t)=N^{-1}\int_{|{\mathcal H}|>0.5{|\mathcal H}|_{max}} |{\mathcal
H}|\,dxdy$ where $N$ is the temporal mean of the integral. The
intensity $I(t)$ exhibits substantial intermittency reflecting the
growth and decay of bursts (thick solid line), while in case $A$ (thin
solid line) the intensity fluctuates relatively little.

\begin{figure}
\epsfxsize=5cm {\epsfbox{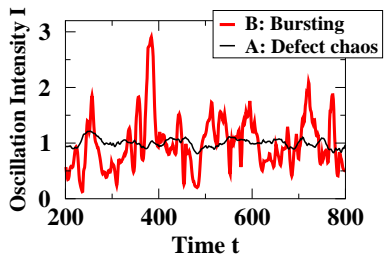}}
\caption{Normalized oscillation intensity $I(t)$ (see text). 
Case $A$ (thin black line) and case $B$ (thick red line).}
\LB{f:burst-t}
\end{figure}

To provide insight into the complex dynamics obtained in our numerical
experiments we make use of the fact that the whirling hexagons arise
from a Hopf bifurcation. It is therefore natural to expect that in the
supercritical case the dynamics can be captured by the CGL for the
oscillation amplitude,
\begin{eqnarray}
\partial_T{\cal H}&=&\mu_2{\cal H}+d\nabla^2{\cal H}-c{\cal H}|{\cal H}|^2. 
\label{e:cgl} 
\end{eqnarray}
We determine the linear coefficients $\mu_2$ and $d\equiv d_r+id_i$
using the Galerkin stability computations and extract the nonlinear
coefficient $c\equiv c_r+ic_i$ from the dependence of the oscillation
frequency on the oscillation amplitude. We find $d=1.3+0.051i$ and
$c=0.95+9.16i$.
Consistent with predictions based on a theory for small-amplitude
hexagons \cite{EcRi00a,EcRi00b}, these values position the system in
the Benjamin-Feir-stable regime ($1+d_ic_i/d_rc_r>0$) in which, however,
defect chaos persists due to the wavenumber selection by the spiral
defects \cite{ChMa96,ArKr02}. Indeed, we find in the NS-simulations
that spatially homogeneous  oscillations are stable. To reach the
chaotic attractor shown in Fig.\ref{f:dis-super} we employ initial
conditions in which the oscillations are not synchronized across the
system.   

To capture the bursting dynamics found in the subcritical case $B$ the CGL
(\ref{e:cgl}) has to be extended to include a quintic term,
$-g|{\mathcal H}|^4{\mathcal H}$. Proceeding as in case $A$, 
we determine the coefficients of the resulting quintic CGL and simulate it in 
two dimensions. As shown in Fig.\ref{f:snap-burst-CGL}a,  which depicts a snapshot of the
magnitude $|{\mathcal H}|$, we obtain quite
similar bursts. Based on this qualitative agreement
we use the CGL to gain further insight into the bursting
found in the NS-simulations.

\begin{figure}
\epsfxsize=3.9cm {\epsfbox{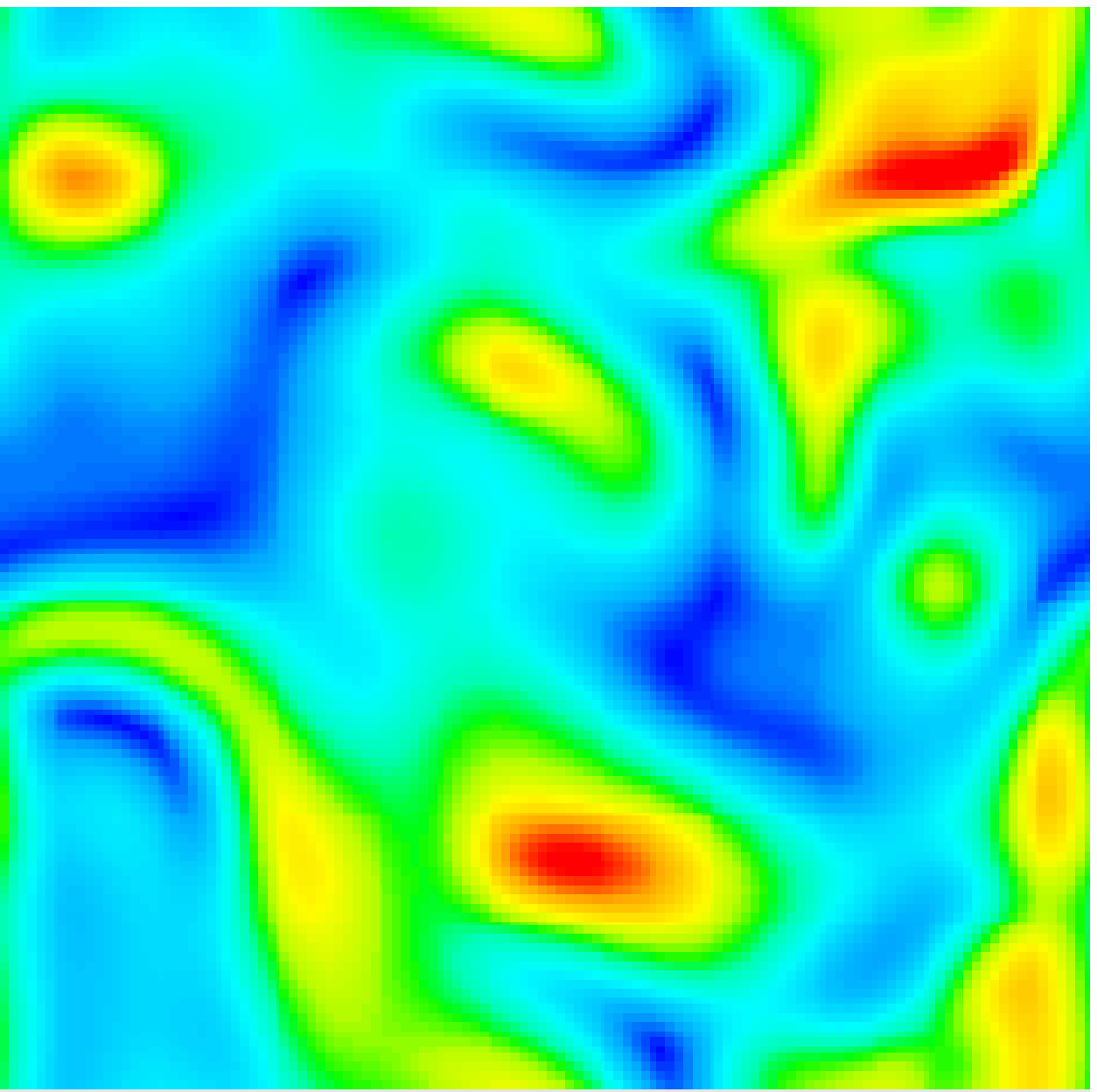}}
\epsfxsize=4.0cm {\epsfbox{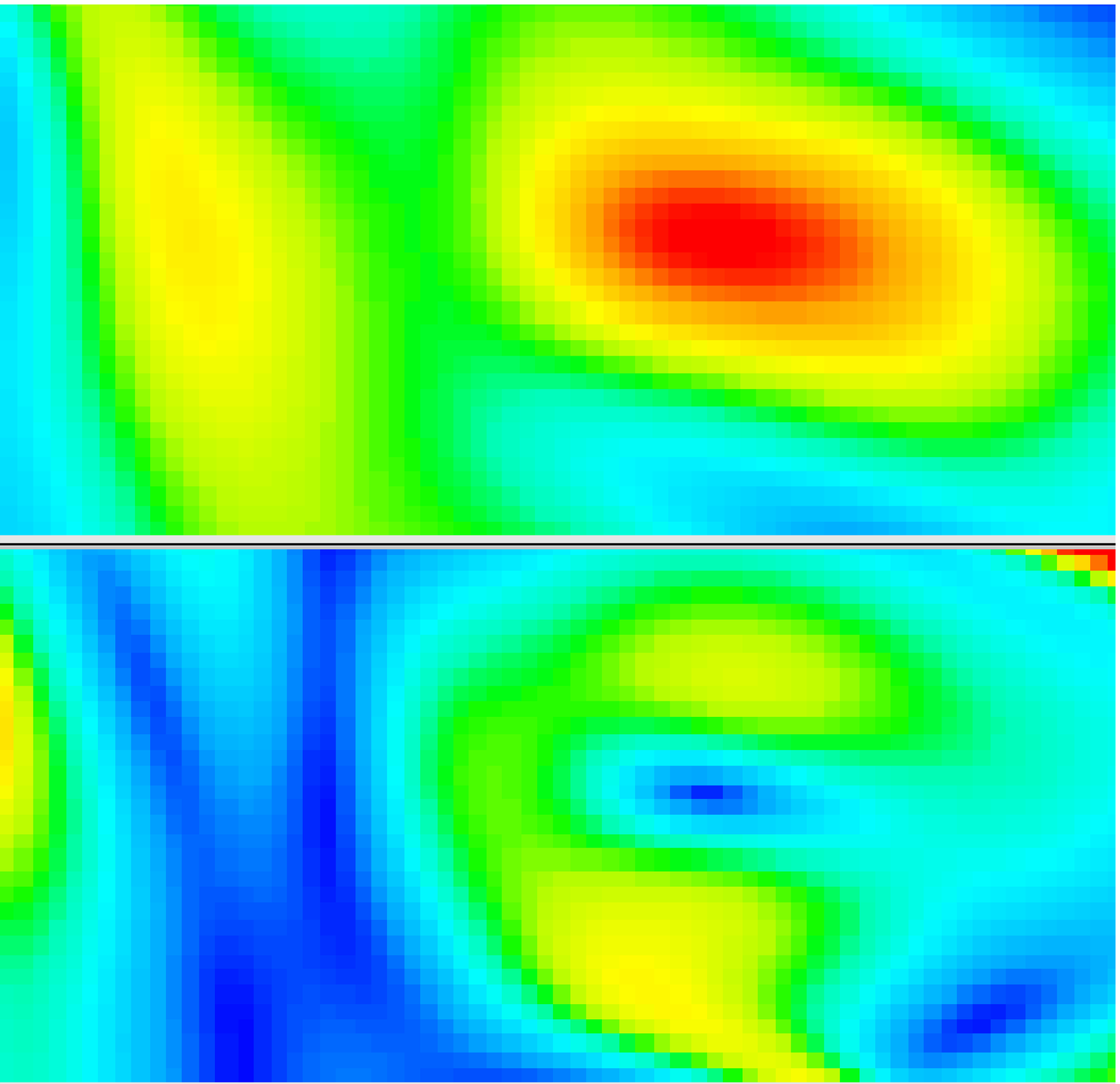}}

\caption{Bursting state obtained with quintic
CGL for parameters corresponding to Fig.\ref{f:snap-burst-NS},
$\mu_2=0.064$, $d=1.90+0.033i$, $c=-1.10+7.19i$, $g=3.61+1.46i$,
except for the larger system size, $L=100$.
a) magnitude $|{\mathcal H}|$ of oscillation, 
b) enlarged portion from a): magnitude $|{\mathcal H}|$ (top) and
associated wavenumber $|{\bf k}|$ of oscillation (bottom).}
\LB{f:snap-burst-CGL}
\end{figure}

The burst mechanism can be elucidated by considering the local
gradient of the phase of ${\mathcal H}\equiv {\mathcal
R}e^{i\psi}$, i.e. the wavevector ${\bf k}\equiv \nabla \psi$ of
the oscillations. Note that ${\bf k}$ is not related to the
wavevectors making up the hexagon pattern. Gradients in
$|{\mathcal H}|$ induce significant differential phase winding
due to the strong amplitude dependence of the oscillation
frequency ($|c_i/c_r|\gg 1$). This leads to a buildup of $|{\bf
k}|$ at the perimeter of the bursts while $|{\bf k}|$ remains
small at their cores, as illustrated in
Fig.\ref{f:snap-burst-CGL}b, which shows an enlargement of a
burst (top) and the associated $|{\bf k}|$ (bottom). The
increased  wavenumber ${\bf k}$ enhances the dissipation via the
diffusion term in (\ref{e:cgl}) and eventually leads to a
collapse of the oscillation amplitude. This mechanism has been studied previously in one dimension \cite{HoSt72,KoGl90,KaKu94,BrSp83,ScKr91,CoKr04}. It has been
shown to underly the dispersive chaos observed in binary-mixture
convection \cite{KoGl90,KaKu94} and, interestingly, can be
strong enough to avoid blow-up even if there is no saturating
nonlinear term at all ($c_r=g_r=0$) \cite{HoSt72,BrSp83,ScKr91}.
The collapse can also be interpreted in terms of two colliding
fronts that connect the steady base state with the nonlinear oscillatory
state. Since these fronts select a non-vanishing wavenumber for
the nonlinear state, the base state can invade the nonlinear
state even above the Hopf bifurcation (`retracting fronts')
\cite{CoKr04}. Note that the collapse of the bursts described here 
is not due to a break-down of the underlying hexagonal structure
\cite{SaKo93,DaWi03}.

Assessing the validity of the CGL (\ref{e:cgl}) and its quintic
extension it is clear that for such a {\it secondary} Hopf
bifurcation a systematic treatment must also include the
coupling of the oscillation mode to slow, long-wave deformations
of the hexagon lattice \footnote{It should be kept in mind that
in contrast to the cubic CGL (\ref{e:cgl}) the quintic equation
can be derived rigorously from the underlying equations only if
both $c_r$ and $c_i$ are small. While this is generically not
the case, it is useful to discuss the quintic CGL as a model
equation.}. They are characterized by gradients in the phase
vector $\vec{\bm
\phi}\equiv\left(\phi_1,(\phi_2-\phi_3)/\sqrt{3}\right)$
\cite{EcRi00a}. In particular, local deformations of the hexagon
lattice enter (\ref{e:cgl}) through an additional term $\sim
{\mathcal H}\nabla \cdot \vec{\bm \phi}$, which captures the
dependence of the growth rate $\sigma$ of the oscillations on
local compressions of the lattice. Conversely, gradients in
${\mathcal H}$ drive $\vec{\bm \phi}$ \cite{EcRi00a}. As a
careful analysis of the NS-simulations shows, they lead to a
decrease of $\nabla \cdot \vec{\bm \phi}$ inside the burst, as
shown in Fig.\ref{f:hopf-sub}a. This has the same effect on
$\sigma$ as a decrease in the wavenumber of the hexagons,
$q_c\rightarrow q_c+\Delta q$ with  $\Delta q=\nabla \cdot
\vec{\bm \phi}/2$, and therefore results in an increase in
$\sigma$ (see Fig.\ref{f:hopf-sub}b), providing a positive
feed-back that enhances the bursting activity compared to the
quintic CGL.

\begin{figure}
\epsfxsize=4.cm {\epsfbox{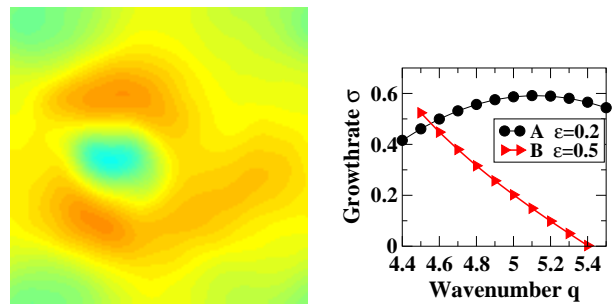}}
\hs{.2cm}
\epsfxsize=3.6cm {\epsfbox{growthrates.supsub.eps}}
\caption{
a) $\nabla \cdot \vec{\bm \phi}$ generated by burst shown in
Fig.\ref{f:snap-burst-NS}b.
b) Growth rate $\sigma$ of the oscillatory mode as a function 
of the wavenumber $q=q_c+\Delta q$.}
\LB{f:hopf-sub}
\end{figure}

Of course, the coupling of the oscillations to the underlying
lattice exists also in the supercritical case shown in
Fig.\ref{f:dis-super}. Consistent with the weakly nonlinear
description of rotating hexagons \cite{EcRi00a,EcRi00b}, we
find, however, that in this regime the growth rate $\sigma$
depends only weakly on the local wavenumber of the hexagons (see
Fig.\ref{f:hopf-sub}b) and varies by less than 10\% across the
system. We therefore suggest that the classic cubic CGL should
describe whirling hexagons very well in the supercritical case
$A$. 

In conclusion, we have numerically investigated whirling hexagons that
arise in rotating NB-convection and have identified two different
regimes. In the weakly NB case the oscillations typically exhibit
defect chaos and our analysis suggests that this state should be well
described by the cubic CGL. Rotating NB-convection would then
represent one of only a few experimentally realizable physical systems  
\cite{OuFl96} in which at least one of the complex states of the two-dimensional CGL is
accessible. 

For stronger NB effects we found that the Hopf bifurcation becomes
subcritical and typical states exhibit bursts. Such bursts and the
related retracting fronts have been discussed in some detail
previously in one dimension
\cite{KoGl90,KaKu94,HoSt72,BrSp83,ScKr91,CoKr04}. The intermittent bursting behavior, however,  has not been studied in detail yet. In two dimensions even less is known.
For instance, the conditions for the persistence of bursting when
stable steady and whirling hexagons compete remain to be understood. Preliminary
simulations indicate that for $\epsilon=0.7$ the bursting persists but
not for $\epsilon=0.9$ (cf. Fig.\ref{f:snap-burst-NS}a).

Since the oscillations arise in a secondary bifurcation the
oscillatory mode is, in principle, coupled to the deformations
of the hexagon lattice. While in the supercritical case our
results indicate that this effect should be small, it is more
significant in the subcritical case, where deformations of the
lattice modify the growth rates of the oscillatory mode
substantially (cf. Fig.\ref{f:hopf-sub}b). Compared to the
bursting behavior of the quintic CGL alone, this leads to an
enrichment of the scenario, which requires further
investigations.

In our simulations we have taken great care to obtain defect-free
hexagonal lattices. Since boundaries tend to introduce defects we have
performed some runs in which circular ramps in the Rayleigh number
mimic a circular container. By stabilizing the hexagon pattern near
the boundary through an additional patterned volume heating we were
able to recover the defect chaos and the bursting state. Conversely,
it would be interesting to allow the underlying lattice to be
disordered. In that case new phenomena may arise from the interaction
of the penta-hepta defects with the whirling mode and significantly
more complex states may be found (cf. \cite{YoRi03b}).

While our numerical experiments have been very productive in
identifying and exploring various complex states in rotating
NB-convection, current computational limitations do not allow us to
use the NS-simulations to investigate in detail the statistical
properties of these states. For instance, for the defect chaos we
cannot address the possibility of deviations from the squared Poisson
distribution for the defect statistics  \cite{HuRi04} or the expected
transition to glassy states with exceedingly slow dynamics
\cite{BrAr03} as the wavenumber of the underlying hexagon pattern is
changed \cite{EcRi00a}. Similarly, investigations of the ramifications
of the hexagonal anisotropy due to the underlying lattice or of the
broken chiral symmetry may require system sizes that are still beyond
current computational capabilities. Questions like these can so far
only be investigated in experiments. 

Two of us (HR, WP) dedicate this paper to the memory of our good
friend Lorenz Kramer.

We gratefully acknowledge support from the Department of Energy
(DE-FG02-92ER14303).

\bibliography{journal}

\end{document}